\documentclass[aps,pra,reprint,amsmath,amssymb,groupedaddress,superscriptaddress,floatfix]{revtex4-2}
\usepackage{graphicx}
\usepackage{bm}
\usepackage{booktabs}
\usepackage{array}
\usepackage{makecell}
\usepackage{microtype}
\usepackage[colorlinks,linkcolor=blue,anchorcolor=blue,citecolor=blue,urlcolor=blue]{hyperref}

\newcommand{\ecl}{\epsilon_{\mathrm{clean}}}
\newcommand{\ep}{\epsilon_p}
\newcommand{\dn}{\delta_{\mathrm{noise}}}
\newcommand{\aeff}{a_{\mathrm{eff}}}
\newcommand{\Egs}{E_{\mathrm{gs}}}
\newcommand{\Ezero}{E_0}
\newcommand{\Rop}{R_{\mathrm{op}}}
\newcommand{\tr}{\operatorname{Tr}}

\newcommand{\ket}[1]{\lvert #1\rangle}
\newcommand{\bra}[1]{\langle #1\rvert}

\graphicspath{{figures/}}

\begin{document}

\title{Initialization-controlled fault sensitivity and force-balance plateaus in noisy variational quantum optimization}

\author{Longhui Shen}
\affiliation{Basic Science Department, Wuchang Shouyi College, Wuhan 430064, China}

\author{Fangzhou Jin}
\affiliation{Basic Science Department, Wuchang Shouyi College, Wuhan 430064, China}

\author{Bin Guo}
\email{binguo@whut.edu.cn}
\affiliation{Department of Physics, Wuhan University of Technology, Wuhan 430070, China}

\date{\today}

\begin{abstract}
We study deep variational circuits for a periodic Heisenberg chain under local gate depolarization, at depths where noiseless training is coherently exact. Full-width initializations reach low coherent error faster than circuits with confined transverse generators, but incur a larger noise shift and need an order of magnitude more updates for noisy convergence. In many-body-localization-inspired initialization, bounding the initial transverse generator amplitudes keeps the operator support localized and suffices for low-fragility optimization, so longitudinal disorder and Floquet layer repetition are unnecessary. The fault-erasure fraction, the mean fraction of coherent energy removed per fault, sets the effective visibility through a parameter-free exponential model. Full-width circuits start with delocalized operator support, and once coherent energy builds up, each fault erases nearly all of it. Initially, their noise-induced gradient almost uniformly attenuates the coherent gradient, so coherent descent proceeds at the noiseless rate. They then stall on a high-noise force-balance plateau, where finite coherent and noise-induced gradients cancel to first order. At late times all initializations reach the same fault-erasure fraction and noisy-error band, but operator support contracts more slowly, at a rate set by the plateau residence time.
\end{abstract}
\maketitle

\section{Introduction}\label{sec:intro}

Variational quantum eigensolvers (VQEs) combine parameterized state preparation with classical optimization and remain a leading approach to near-term quantum simulation \cite{Peruzzo2014,Cerezo2021,Tilly2022,Bharti2022}. Their reach depends on circuit expressivity and on the competition between gradient trainability and gate-level dissipation \cite{Tilly2022,Bharti2022,Leone2024,Cerezo2025,Larocca2022,Wiersema2020,Takagi2022}. Deep entangling circuits are needed to represent nontrivial many-body ground states, but generic random initializations suffer from barren plateaus in which gradients vanish exponentially with system size \cite{McClean2018,Cerezo2021a,Holmes2022,OrtizMarrero2021,Larocca2025}, and hardware noise further flattens the landscape and limits the useful depth \cite{Wang2021,Mele2026,Endo2021,Singkanipa2025,Yan2025}. Structured initialization is therefore widely used \cite{Zou2025}. Identity-block and layerwise schemes preserve local gradients \cite{Grant2019,Skolik2021,Volkoff2021}, and constrained parameter ranges remove barren plateaus in Hamiltonian variational ans\"atze \cite{Park2024}. Floquet- and many-body-localization (MBL)-inspired circuits suppress entanglement growth and retain trainability by mimicking slow scrambling in localized phases \cite{Cao2025,Yin2025,Bardarson2012,Shen2026,Serbyn2013,Nandkishore2015,Abanin2019}.

A favorable initial gradient, however, does not by itself guarantee successful optimization, and both the full-trajectory dynamics and the structural requirements of these designs remain open. Warm-start analyses separate initial trainability from later convergence \cite{Puig2025}, and parameter-dependent noise resilience, together with error-penalized and truncated search strategies, has so far been established mainly for shallow optimization landscapes \cite{Kirk2026}. How coherent energy relaxation and fault amplification co-evolve over an entire noisy training trajectory is not known. Because the Heisenberg light cone that builds multisite correlations to lower the ideal energy also converts local gate errors into global energy shifts \cite{Lieb1972,Roberts2017,Nahum2018,Keyserlingk2018,Temme2017,Endo2021}, rapid coherent descent could either accelerate noisy convergence or generate the fault sensitivity that later stalls the optimizer.

It is also unclear which structural feature of MBL-inspired initialization controls this noisy evolution. MBL-inspired schemes combine three priors: small transverse rotation angles, strong longitudinal phase disorder, and Floquet layer repetition \cite{Cao2025,Yin2025,Ponte2015,Lazarides2015}. Studies of small-angle and Gaussian initializations \cite{Grant2019,Holmes2022,Park2024}, together with MBL-inspired gradient measurements and VQE experiments on noisy hardware \cite{Cao2025,Yin2025}, show that such circuits train well, but they do not separate the roles of the three priors over a long noisy trajectory. Consequently, it is unknown whether full-width disorder and temporal periodicity are needed at all, or whether a simpler restriction on the generators carries the protection.

We address both questions for a periodic Heisenberg chain ($N=8$, $L\in\{8,16\}$, with an $N=10$ reference case) under idealized local gate-level depolarization. A five-group hierarchical ablation (Groups A--E) removes Floquet layer repetition and longitudinal disorder step by step from the MBL baseline. Each trajectory is decomposed into coherent error $\ecl$ and noise shift $\dn$, and snapshots at seven times record the single-fault energy susceptibility, the operator radius of a propagated fault, and the geometry of coherent and noise-induced gradients. Finally, constrained iso-energetic tangent projections test whether accumulated fragility can be removed post hoc. We obtain four results. (i) In the coherently exact regime ($N=8$, $L=16$), fast early coherent descent and slow noisy convergence coexist: full-width circuits cross 5\% coherent replay error three times faster along noisy trajectories (and about $1.35$ times faster in noiseless training) but need an order of magnitude more updates to reach noisy-error targets. (ii) The low-fragility regime requires only bounded initial transverse generator amplitudes; full-width longitudinal disorder and Floquet layer repetition are unnecessary, and narrowing the longitudinal angles as well (Group E) even shortens late-stage convergence. (iii) The fragility is energetic rather than spatial. It is governed by the fault-erasure fraction $F$, which is already near unity in full-width circuits from the first snapshot with appreciable coherent energy. An independent-fault exponential model in $F$ reproduces the exact noisy energies. It also links the measured gradient geometry---weak early braking in full-width circuits followed by a first-order force-balance plateau---to the evolution of the effective visibility. (iv) The average energy response relaxes to a common value within 600 updates, while the spatial support of an early-fault probe contracts on a slower, plateau-dependent time scale. Post hoc straight-step tangent repair is limited both by the small tangential drive near force balance and by the $\sqrt{\tau}$-bounded second-order departure from the coherent-energy level set.

Section~\ref{sec:model} defines the model, ablation design, and diagnostics. Section~\ref{sec:results} establishes the speed--fragility tradeoff and the ablation result. Section~\ref{sec:mechanism} analyzes the phase-space trajectories, the fault-erasure mechanism, and the force-balance plateau. Section~\ref{sec:projection} presents the tangent-projection experiments, and Sec.~\ref{sec:conclusion} concludes.

\section{Circuit model and diagnostic framework}\label{sec:model}

\subsection{Model Hamiltonian and noisy circuit ansatz}

We consider an even-$N$ periodic spin-$1/2$ chain with the isotropic antiferromagnetic Heisenberg Hamiltonian \cite{Bethe1931},
\begin{equation}\label{eq:H}
H=\sum_{\langle i,j\rangle\in\mathcal E}(X_iX_j+Y_iY_j+Z_iZ_j),
\end{equation}
and define the cost function as the energy density
\begin{equation}\label{eq:cp}
c_p(\bm\theta)=\frac{\tr[H\rho_p(\bm\theta)]}{N},
\end{equation}
where $X_i, Y_i, Z_i$ are Pauli operators, $\mathcal{E}$ denotes the nearest-neighbor ring bonds, and $\rho_p(\bm\theta)$ is the noisy output state. Exact diagonalization gives the ground-state energy $\Egs<0$ ($\Egs=-14.6044$ at $N=8$). Starting from $\ket{0}^{\otimes N}$, the circuit applies $L$ parameterized layers \cite{Wecker2015} and a final N\'eel decoder ($R_x(\pi)$ on odd sites). Each layer sequentially applies single-site $R_x$ and $R_z$ rotations, followed by two-qubit $R_{xx}$, $R_{yy}$, and $R_{zz}$ blocks across $\mathcal{E}$, with $R_\alpha(\theta)=\exp(-i\theta P_\alpha/2)$ ($P_\alpha \in \{X_i, Z_i, X_iX_j, Y_iY_j, Z_iZ_j\}$). Within each two-qubit block, gates act on odd bonds $(1,2),(3,4),\dots,(N-1,N)$ then even bonds $(2,3),(4,5),\dots,(N,1)$, yielding $P=5NL$ independent angles in $[-\pi,\pi)$. The longitudinal generators $\{Z_i, Z_iZ_j\}$ commute and are diagonal in the computational basis, whereas the transverse generators $\{X_i, X_iX_j, Y_iY_j\}$ drive off-diagonal transitions and operator spreading. Because the N\'eel flips are deferred to the decoder, $\ket{0}^{\otimes N}$ is a joint eigenstate of all longitudinal generators; longitudinal gates thus act nontrivially only on superpositions created by the magnetization-nonconserving transverse gates. At $N=8$ and $L=16$, $P=640$ exceeds the real dimension $2^{N+1}-2=510$ of the projective Hilbert space, a necessary condition for overparameterization in the full state space \cite{Larocca2023}, and noiseless full-width training indeed reaches $\ecl\sim10^{-9}\%$, as shown in Fig.~\ref{fig:depth}(c). Neither the $L=8$ circuit ($P=320$) nor the $N=10$, $L=16$ reference case ($P=800<2^{11}-2=2046$) satisfies this bound.

Every elementary gate (including decoder rotations) is followed by a local depolarizing channel \cite{Cai2023} on each participating qubit,
\begin{equation}\label{eq:depol}
\mathcal E_p(\rho)=(1-p)\rho+\frac{p}{3}(X\rho X+Y\rho Y+Z\rho Z),
\end{equation}
acting as $\mathcal E_p^{(i)}\otimes\mathcal E_p^{(j)}$ after two-qubit gates on $(i,j)$. Each layer introduces $2N+6N=8N$ single-qubit fault locations and the decoder adds $N/2$, for a total of $K=8NL+N/2$ ($K=516$ and $1028$ for $L=8$ and $16$ at $N=8$). Both the noiseless state $\ket{\psi_0(\bm\theta)}$, with noiseless replay energy $\Ezero(\bm\theta)=\bra{\psi_0}H\ket{\psi_0}$, and the noisy density matrix $\rho_p(\bm\theta)$ \cite{Fontana2025} are simulated exactly in double precision. Exact gradients from reverse-mode automatic differentiation \cite{Jones2020,Schuld2019} drive Adam updates \cite{Kingma2015} ($\eta=0.03$, $\beta_1=0.9$, $\beta_2=0.999$, $\epsilon_{\mathrm{Adam}}=10^{-8}$). Throughout, an update denotes one exact-gradient Adam step, not a measurement shot. Reported values are seed medians (with interquartile ranges, IQRs, where noted), except for the five-seed mean operator radii $\Rop$ at $N=8$.

\subsection{Hierarchical ablation design}

Table~\ref{tab:groups} summarizes the five initialization groups. Groups A--D form a $2\times2$ factorial design crossing initial transverse generator width with layer repetition at full-width longitudinal angles $[-\pi,\pi)$; the pairs A--B and C--D isolate the effect of layer repetition at confined and full transverse widths, respectively. Group E additionally narrows the longitudinal angles to $[-0.2,0.2]$ with independent layers, so that the comparison of B and E tests whether broad longitudinal disorder is still required once the transverse angles are confined. In the repeated-layer groups, layer repetition is imposed only on the initial parameters $\bm\theta(0)$; all $P$ parameters are subsequently updated independently.

\begin{table*}[t]
\caption{Five-group hierarchical ablation. Group A is the Floquet--MBL baseline; Group B removes layer repetition; Group E further narrows the longitudinal angle range. Groups C and D serve as full-width controls. Initial angles are drawn uniformly from the specified intervals.}
\label{tab:groups}

\centering
\small
\begin{tabular*}{\textwidth}{@{\extracolsep{\fill}}lcccl@{}}
\toprule
Group & Transverse ($R_x,R_{xx},R_{yy}$) & Longitudinal ($R_z,R_{zz}$) & Initial layers & Role \\
\midrule
A & $[-0.2,0.2]$ & $[-\pi,\pi)$ & Repeated
  & Floquet--MBL baseline \\
B & $[-0.2,0.2]$ & $[-\pi,\pi)$ & Independent
  & Confined transverse generators, independent layers \\
C & $[-\pi,\pi)$ & $[-\pi,\pi)$ & Repeated
  & Full-width control with layer repetition \\
D & $[-\pi,\pi)$ & $[-\pi,\pi)$ & Independent
  & Full-width random baseline \\
E & $[-0.2,0.2]$ & $[-0.2,0.2]$ & Independent
  & All angles confined (narrow longitudinal range) \\
\bottomrule
\end{tabular*}
\end{table*}

Our primary simulations cover $N=8$, $L\in\{8,16\}$, and $p\in\{0,10^{-3}\}$ with 20 seeds per group. Noiseless and $L=8$ runs last 300 updates, whereas the noisy runs at $L=16$, $p=10^{-3}$ (Groups A--E) continue for 600 updates and are evaluated at $t=300$ and $t=600$. Microscopic diagnostics are computed for seeds 0--4 for Groups A--D at seven snapshots $t\in\{0,20,50,100,200,300,600\}$ (140 configurations), from which tangent-space projections branch at $t=100$ and $t=300$. Supplementary runs comprise noisy training at a smaller learning rate $\eta=0.01$ (five seeds per group, 600 updates), noiseless reachability runs for the $N=10$, $L=16$ reference case (five seeds per group, up to 600 updates), and 50-seed initial-gradient ensembles for shot-resolvability estimates.

\subsection{Phase-space coordinates and reference states}

At each update, we evaluate both the noisy energy $E_p(\bm\theta)$ and the noiseless replay energy $\Ezero(\bm\theta)$ to decompose the relative noisy energy error as
\begin{equation}\label{eq:decomp}
\ep\equiv\frac{E_p-\Egs}{|\Egs|}=\ecl+\dn,
\end{equation}
with
\begin{equation}\label{eq:definitions}
\ecl=\frac{\Ezero-\Egs}{|\Egs|},\qquad \dn=\frac{E_p-\Ezero}{|\Egs|},
\end{equation}
Here $\ecl\ge0$ is the relative coherent energy error and $\dn$ is the signed noise-induced energy shift at fixed $\bm\theta$ (which can be negative when $\Ezero>0$).

Because $\tr\,H=0$, a state-independent global depolarizing channel \cite{StilckFranca2021} would contract the energy uniformly, $E_p=a\Ezero$ with a constant visibility $a$, so that trajectories would lie on the straight line
\begin{equation}\label{eq:null}
\dn=(1-a)(1-\ecl).
\end{equation}
For a general circuit, the departure from this line is measured by the instantaneous effective visibility
\begin{equation}\label{eq:aeff}
\aeff(\bm\theta)\equiv\frac{E_p}{\Ezero}
=1-\frac{\dn}{1-\ecl}.
\end{equation}
The effective visibility is defined for $\Ezero\neq0$ and lies in $[0,1]$ whenever $\Ezero\le E_p\le0$, which holds for all analyzed points with $\ecl<0.9$. Equation~\eqref{eq:null} serves as a uniform-attenuation null model; Sec.~\ref{sec:mechanism} fits it to all 601 points of each seed and uses it for the reference lines $a=0.40$ and $0.78$. We define a 20-update interval as an iso-energetic denoising window if $|\Delta\ecl|\le0.1$ percentage points (pp) while $\Delta\dn\le-0.1$~pp.

Two shallow states calibrate the error scale at $N=8$ and $p=10^{-3}$. The N\'eel product state prepared by the decoder alone yields $\ep=45.29\%$. A stronger benchmark is the four-singlet dimer product $\ket{\Psi_{\mathrm{dim}}}=\ket{\psi^-}_{12}\ket{\psi^-}_{34}\ket{\psi^-}_{56}\ket{\psi^-}_{78}$, which has ideal energy $E_{\mathrm{dim}}=-12$ ($\ecl=17.83\%$). Each singlet is prepared from $\ket{00}$ by $R_x^{(i)}(\pi)$, $R_{xx}^{(i,j)}(\pi/2)$, and $R_z^{(i)}(\pi/2)$, giving 16 fault locations in total under Eq.~\eqref{eq:depol}. Because single-site marginals of each noisy pair remain maximally mixed, inter-pair bonds contribute zero energy, yielding the exact noisy energy
\begin{equation}\label{eq:dimer}
E_{\mathrm{dim},p}=-4(2\lambda^4+\lambda^3),
\end{equation}
with $\lambda=1-4p/3$, corresponding to $\ep=18.23\%$ at $p=10^{-3}$.

Convergence speed is quantified by first-passage times. For a target $\epsilon_\star\in\{25\%,30\%,35\%\}$, the first-passage time is the first update with $\ep\le\epsilon_\star$, and $t_{50}$ is the update at which the cumulative hitting fraction across seeds reaches 50\%, with unhit runs retained as right-censored. Reported endpoint values correspond to the final update ($t=300$ or $600$). The $N=10$ noiseless reference runs stop upon reaching $\ecl\le5\%$ or at $t=600$.

\subsection{Microscopic probes and tangent-projection protocol}\label{sec:probes}

We quantify the energetic sensitivity to isolated faults by the zero-noise susceptibility, the linear response of the energy to the replay noise rate $q$ of the channel in Eq.~\eqref{eq:depol} \cite{Dalton2024}:
\begin{equation}\label{eq:chi}
\chi=\left.\frac{\partial E_q}{\partial q}\right|_{q=0}
=\sum_{s=1}^{K}\Delta E_s,
\end{equation}
with the local mean Pauli-fault response
\begin{equation}\label{eq:deltaEs}
\Delta E_s=\frac13\sum_{Q\in\{X,Y,Z\}}(E_{s,Q}-\Ezero),
\end{equation}
where $E_{s,Q}$ is the noiseless energy with a single Pauli fault $Q$ inserted at location $s$. We estimate $\chi$ from a uniform sample $\mathcal{S}$ of $m=24$ locations drawn without replacement,
\begin{equation}\label{eq:chihat}
\widehat\chi=\frac Km\sum_{s\in\mathcal S}\Delta E_s,\qquad m=24,
\end{equation}
which is an unbiased estimator with finite-population standard error $K[(1-m/K)s_{\Delta E}^2/m]^{1/2}$. For $\Ezero<0$, we define the fault-erasure fraction
\begin{equation}\label{eq:F}
F\equiv\frac{\chi}{K|\Ezero|}
=\frac1K\sum_{s=1}^{K}f_s,\qquad
f_s=\frac{\Delta E_s}{|\Ezero|},
\end{equation}
which measures the mean fraction of coherent energy removed per fault (as signed responses, $f_s$ and $F$ are not restricted to $[0,1]$). A fault that drives the state to the zero-energy white-noise baseline gives $f_s=1$, so $F\simeq1$ marks the energetic white-noise limit of a scrambling circuit \cite{Dalzell2024}. Because $\chi$ and $F$ are evaluated at zero noise, they can be compared directly with exact noisy energies at the training rate $p$ and replay noise rates $q\in\{10^{-5},10^{-4},10^{-3}\}$.

As a spatial diagnostic, we follow the spreading of a single early fault through the full circuit depth. We inject a Pauli fault $V\in\{X,Y,Z\}$ after the first-layer $R_x$ gate on site $j_0=4$ and track its Heisenberg evolution $O(\ell)=U_\ell VU_\ell^\dagger$ ($\tr(O^\dagger O)=2^N$) through the noiseless suffix unitary $U_\ell$ up to layer $\ell$ ($\ell=0$ at injection, $\ell=1,\dots,16$ after subsequent circuit layers, and $\ell=17$ after the decoder). The spatial support is characterized by the site-resolved weights
\begin{equation}\label{eq:w}
w_j=\frac{\displaystyle\sum_{Q\in\{X,Y,Z\}}
 \lVert[O,Q_j]\rVert_F^2}{8\,2^N},
\end{equation}
and the operator radius
\begin{equation}\label{eq:radius}
\Rop=\left[\frac{\sum_j d(j,j_0)^2w_j}{\sum_j w_j}\right]^{1/2},
\end{equation}
where $d(j,j_0)$ is the ring distance and $w_j$ is an out-of-time-order-correlator weight \cite{Swingle2016} averaged over $V\in\{X,Y,Z\}$ that vanishes if and only if $O$ acts trivially on site $j$. A uniform spatial profile yields the complete-delocalization reference $R_{\mathrm{op}}^{\mathrm{unif}}=[(N^2+2)/12]^{1/2}$ ($2.3452$ at $N=8$ and $2.9155$ at $N=10$). Because single-site decoder rotations preserve $w_j$, $\Rop(16)=\Rop(17)$. At the circuit output ($\ell=17$), a fault $Q$ at a general location $s$ with suffix unitary $S_s$ acts on $\ket{\psi_0}$ as $W_{s,Q}=S_sQS_s^\dagger$, producing the energy shift
\begin{align}\label{eq:commutator}
\Delta E_{s,Q}&=\bra{\psi_0}W_{s,Q}^\dagger H W_{s,Q}-H\ket{\psi_0}\nonumber\\
&=\bra{\psi_0}W_{s,Q}^\dagger[H,W_{s,Q}]\ket{\psi_0}.
\end{align}
$\Rop$ therefore measures the state-independent spatial spreading of an early fault across the full circuit depth, whereas $\Delta E_{s,Q}$ (and hence $F$) depends on the commutator expectation in the instantaneous state $\ket{\psi_0}$.

To characterize the interplay between coherent optimization and noise amplification in parameter space, we decompose the total gradient as $\nabla_{\bm\theta}\ep=\bm g_0+\bm g_\delta$, with $\bm g_0=\nabla_{\bm\theta}\ecl$ and $\bm g_\delta=\nabla_{\bm\theta}\dn$, and define
\begin{align}\label{eq:geometry}
\cos\phi&=\frac{\bm g_0^\top\bm g_\delta}{\lVert\bm g_0\rVert\lVert\bm g_\delta\rVert},\\
r&=\frac{\lVert\bm g_\delta\rVert}{\lVert\bm g_0\rVert},\\
\kappa&=\frac{\lVert\bm g_0+\bm g_\delta\rVert}{\lVert\bm g_0\rVert+\lVert\bm g_\delta\rVert},\label{eq:kappa}\\
\cos\psi&=\frac{1+r\cos\phi}{\sqrt{1+r^2+2r\cos\phi}},
\end{align}
where $\kappa=[(1+r^2+2r\cos\phi)/(1+r)^2]^{1/2}$ quantifies vector cancellation and $\cos\psi$ measures the alignment between the total gradient and $\bm g_0$. First-order force balance corresponds to $\cos\phi\to-1$, $r\to1$, and $\kappa\to0$ at finite gradient norms $\lVert\bm g_0\rVert,\lVert\bm g_\delta\rVert>0$.

To test whether accumulated noise can be suppressed at fixed coherent error, we project the steepest noise-descent direction $-\bm g_\delta$ onto the tangent space of the constant-$\ecl$ manifold \cite{Boumal2023}:
\begin{equation}\label{eq:d}
\bm d=-\bm g_\delta+
\bm g_0\frac{\bm g_0^\top\bm g_\delta}{\lVert\bm g_0\rVert^2}.
\end{equation}
From each snapshot $\bm\theta_{\mathrm{anchor}}$ ($t=100$ or $300$), we perform 50 projected steps. At step $k$, a trial update $\bm\theta_k+\alpha\bm d$ (with $\lVert\bm d\rVert$ capped at $0.3$) is tested via backtracking line search $\alpha=2^{-n}$ ($n=0,\dots,15$). A step is accepted if it satisfies the Armijo condition \cite{Armijo1966} $\dn(\bm\theta_k+\alpha\bm d)\le\dn(\bm\theta_k)+10^{-4}\alpha\,\bm g_\delta^\top\bm d$ together with one of three coherence constraints relative to $\bm\theta_{\mathrm{anchor}}$: $|\ecl(\bm\theta_{k+1})-\ecl(\bm\theta_{\mathrm{anchor}})|\le\tau$ with $\tau=0.2$~pp (tolerant) or $0.02$~pp (tight), or $\ecl(\bm\theta_{k+1})\le\ecl(\bm\theta_{\mathrm{anchor}})+10^{-12}$ (strict nondegradation). If all 16 trial lengths are rejected, $\bm\theta_k$ remains unchanged. Each 50-step branch consumes 100 full-gradient evaluations ($\bm g_0$ and $\bm g_\delta$) plus line-search energy evaluations, and is compared with an unconstrained Adam restart with the same budget of 100 gradient evaluations.

Finally, the single-gate resolvability of the initial gradients reported in Sec.~\ref{sec:robust} is estimated with the parameter-shift rule \cite{Wierichs2022,Mari2021}. Splitting $M$ measurement shots per derivative equally between the two shifted circuits ($\theta_\mu\pm\pi/2$) and uniformly across the $3N$ Pauli bond terms ($M/(6N)$ shots per term) yields the estimator
\begin{equation}\label{eq:shotg}
\widehat g_\mu=\frac{1}{2N}\sum_{a=1}^{3N}
(\widehat P_{a,+}-\widehat P_{a,-}),
\end{equation}
with variance
\begin{equation}\label{eq:shotsigma}
\sigma_{\widehat g_\mu}^{2}
=\frac{3}{2NM}\sum_{a=1}^{3N}
\left[(1-\langle P_a\rangle_+^2)+(1-\langle P_a\rangle_-^2)\right].
\end{equation}
A gate derivative at $t=0$ is counted as resolved if $|g_\mu|/\sigma_{\widehat g_\mu}\geq2$.

\section{Speed--fragility tradeoff and prior ablation}\label{sec:results}

\subsection{Separation of coherent and noisy relaxation times}\label{sec:separation}

\begin{figure*}[t]
\includegraphics[width=\textwidth,clip]{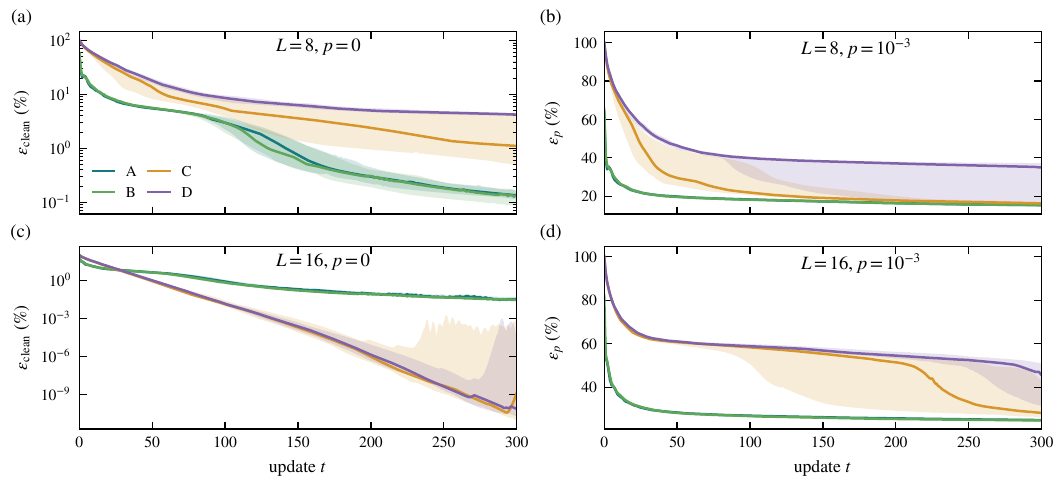}
\caption{Depth and noise dependence of Groups A--D over 300 updates at $N=8$ (20 seeds per condition; medians and IQRs): (a) $L=8$, $p=0$; (b) $L=8$, $p=10^{-3}$; (c) $L=16$, $p=0$; (d) $L=16$, $p=10^{-3}$. Noiseless panels show $\ecl$ on a logarithmic axis.}\label{fig:depth}
\end{figure*}

As shown in Fig.~\ref{fig:depth}, circuit depths $L=8$ and $L=16$ probe distinct optimization regimes. At $L=8$ ($K=516$) and $p=10^{-3}$, Groups A--C reach noisy errors of $\ep=15.4\%$--$16.3\%$ within 300 updates (Group D: $35.0\%$), outperforming both the noisy N\'eel state ($45.29\%$) and the singlet-dimer benchmark ($18.23\%$), whereas noiseless $L=8$ circuits saturate at $\ecl=0.13\%$--$4.2\%$. Doubling the depth to $L=16$ renders the full-width circuits coherently exact ($\ecl\sim10^{-9}\%$ at $p=0$) while doubling the fault count to $K=1028$. Consequently, the 600-update noisy endpoints at $L=16$ ($\ep=23.6\%$--$24.6\%$) lie above the shallow singlet-dimer benchmark, though well below the N\'eel reference. Using the independent-fault relation $\dn\simeq1-e^{-pKF}$ derived in Sec.~\ref{sec:mechanism} from Eq.~\eqref{eq:expvis}, the measured $L=16$ endpoint value $F\simeq0.26$ (and the effective $F\simeq0.30$ inverted from the $L=8$ endpoint medians) accounts for the growth of residual noise shift from ${\sim}14\%$ to ${\sim}23\%$ as $Kp$ doubles. Thus, $L=16$ isolates the regime where a coherently exact circuit contends with doubled fault exposure, giving rise to the two-stage relaxation dynamics analyzed below.

As listed in Table~\ref{tab:hitclean}, coherent convergence exhibits the opposite ordering across initializations. In noiseless training ($p=0$), full-width circuits (Groups C and D) reach $\ecl\le5\%$ in 30.5--32 updates and $\ecl\le1\%$ in 49.5--51 updates, compared with 41.5--43.5 and 90.5--94.5 updates for confined circuits (Groups A and B). Full-width circuits initialize near $\Ezero\simeq0$ with all noncommuting couplings strongly active, whereas confined circuits start near the N\'eel product state ($\ecl=54\%$ and $68\%$ at $t=0$) and build entanglement gradually through weak transverse gates. Along noisy trajectories ($p=10^{-3}$), the noiseless replay energy of Groups C and D still crosses $\ecl=5\%$ at 31 and 32 updates---matching their noiseless rate---whereas Groups A, B, and E are retarded to 104, 104, and 80 updates, respectively. Gate noise thus leaves early coherent descent intact in full-width circuits while significantly slowing confined ones, an asymmetry that Sec.~\ref{sec:mechanism} traces to the gradient geometry. Despite this $1.35$-fold noiseless and threefold noisy-replay speedup in early coherent descent, full-width initialization requires an order of magnitude more updates to converge in noisy energy.

\begin{table*}[t]
\caption{Coherent-error first-passage times by protocol: independent noiseless training (300 updates), coherent replay along $p=10^{-3}$ training (600 updates), and replay along $\eta=0.01$ training (600 updates). Entries give the median first-passage update among successful seeds and hits/seeds; unhit runs are censored at the budget. Group E was not run at $p=0$ or $\eta=0.01$.}\label{tab:hitclean}
\centering\small
\begin{tabular*}{\textwidth}{@{\extracolsep{\fill}}lccccc@{}}
\toprule
& \multicolumn{2}{c}{Noiseless ($p=0$)} & \multicolumn{2}{c}{Replay ($p=10^{-3}$)} & $\eta=0.01$ \\
\cmidrule(lr){2-3}\cmidrule(lr){4-5}
Group & $5\%$ & $1\%$ & $5\%$ & $1\%$ & $5\%$ \\
\midrule
A & 43.5 (20/20) & 94.5 (20/20) & 104 (20/20) & 519 (3/20) & 190 (5/5) \\
B & 41.5 (20/20) & 90.5 (20/20) & 104 (20/20) & 419 (1/20) & 171 (5/5) \\
C & 30.5 (20/20) & 49.5 (20/20) & 31 (20/20) & 55 (16/20) & 49 (5/5) \\
D & 32 (20/20) & 51 (20/20) & 32 (20/20) & 56 (20/20) & 54 (5/5) \\
E & -- & -- & 80 (20/20) & 490 (8/20) & -- \\
\bottomrule
\end{tabular*}
\end{table*}

\begin{figure*}[t]
\includegraphics[width=2\columnwidth,clip]{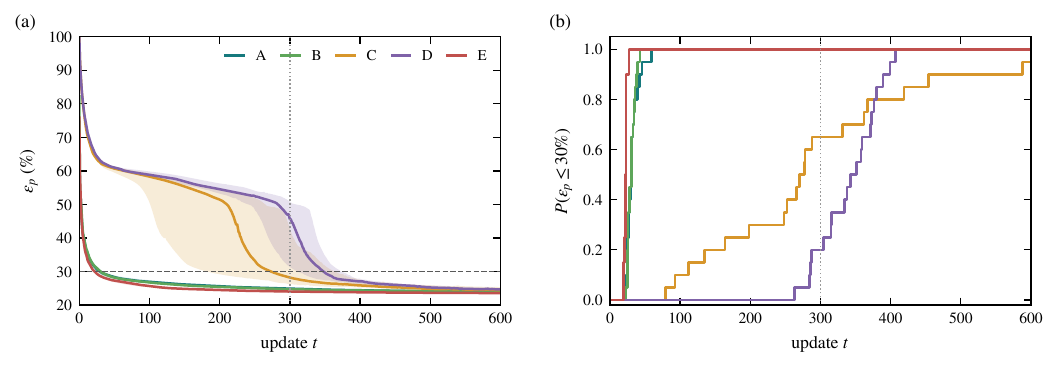}
\caption{Noisy training at $N=8$, $L=16$, $p=10^{-3}$ (20 seeds per group). (a) Median noisy error $\ep$ with interquartile bands; dotted line, $t=300$; dashed line, 30\% target. (b) Fraction of seeds reaching $\ep\leq30\%$ by update $t$.}\label{fig:main}
\end{figure*}

Figure~\ref{fig:main} and Table~\ref{tab:endpoints} summarize the $N=8$, $L=16$, $p=10^{-3}$ trajectories. At $t=300$, median noisy errors span $21.56$~pp across groups: $24.02\%$ (E), $24.72\%$ (B), $24.89\%$ (A), $28.21\%$ (C), and $45.58\%$ (D). By $t=600$, all five groups converge into a narrow band between $23.56\%$ and $24.64\%$ (a $0.65$~pp spread across A--D, with $\ecl\in[1.12\%,1.89\%]$ and $\dn\in[22.01\%,23.01\%]$), while continuing to decrease slowly by $0.025$--$0.136$~pp over the final 20 updates. The large spread at $t=300$ therefore reflects disparate relaxation rates rather than distinct asymptotic noise floors.

\begin{table*}[t]
\caption{Endpoint errors and first-passage statistics at $N=8$, $L=16$, $p=10^{-3}$ (20 seeds per group, 600 updates). Here $t_{50}$ is the first update at which the cumulative hitting fraction reaches 50\% (retaining unhit runs as right-censored). The decomposition $\ep=\ecl+\dn$ holds per seed; group medians are not strictly additive.}\label{tab:endpoints}
\centering\scriptsize
\begin{tabular*}{\textwidth}{@{\extracolsep{\fill}}lccccccc@{}}
\toprule
Group & $\ep(300)$ (\%) & $\ep(600)$ (\%) & $\ecl(600)$ (\%) & $\dn(600)$ (\%) & $25\%$: hits, $t_{50}$ & $30\%$: hits, $t_{50}$ & $35\%$: hits, $t_{50}$ \\
\midrule
A & 24.89 & 24.01 & 1.89 & 22.27 & 20/20, 279 & 20/20, 30 & 20/20, 13 \\
B & 24.72 & 23.99 & 1.71 & 22.01 & 20/20, 254 & 20/20, 31 & 20/20, 13 \\
C & 28.21 & 24.43 & 1.59 & 23.01 & 14/20, 471 & 19/20, 270 & 19/20, 241 \\
D & 45.58 & 24.64 & 1.60 & 22.96 & 14/20, 524 & 20/20, 343 & 20/20, 316 \\
E & 24.02 & 23.56 & 1.12 & 22.43 & 20/20, 130 & 20/20, 22 & 20/20, 10 \\
\bottomrule
\end{tabular*}
\end{table*}

First-passage times $t_{50}$ confirm an order-of-magnitude separation in noisy relaxation speed. For the $\epsilon_\star=30\%$ target, confined groups reach $t_{50}$ in 22 (E), 30 (A), and 31 (B) updates, whereas full-width groups require 270 (C) and 343 (D) updates. The resulting full-width-to-confined $t_{50}$ ratios span $8.7$ (C/B) to $15.6$ (D/E), with 95\% nonparametric bootstrap intervals (5000 resamples) between $6.4$ and $16.9$. Similarly, $t_{50}$ at $\epsilon_\star=35\%$ is 10--13 updates for Groups A, B, and E versus 241--316 for C and D. At $\epsilon_\star=25\%$, near the late-time band, $t_{50}$ is 130--279 versus 471--524 updates, narrowing the ratio to $1.7$--$4.0$ as all trajectories enter slow late-stage denoising. All 20 seeds in Groups A, B, and E cross every threshold, whereas Group C hits $30\%$ and $35\%$ in 19/20 seeds, and both C and D reach $25\%$ in only 14/20 seeds. We term this combination of rapid early coherent descent, a shared late-time noisy energy band, and an order-of-magnitude delay in noisy first-passage times the \emph{speed--fragility tradeoff}.

\subsection{Transverse generator confinement versus disorder and periodicity}

Comparing Groups A, B, and E isolates the structural prior responsible for low-fragility training. All three confined groups follow closely matched low-noise trajectories, reaching $\ep(600)=24.01\%$, $23.99\%$, and $23.56\%$ with $t_{50}(30\%)=30$, 31, and 22 updates, respectively. Removing Floquet layer repetition at confined transverse width (Group B vs.\ A) leaves first-passage times and endpoint noise shifts ($22.01\%$ vs.\ $22.27\%$) virtually unchanged. Furthermore, narrowing the initial longitudinal angle range from $[-\pi,\pi)$ to $[-0.2,0.2]$ (Group E vs.\ B) preserves the low-fragility regime while yielding the lowest endpoint coherent error ($1.12\%$ vs.\ $1.71\%$--$1.89\%$) and roughly halving the $25\%$ first-passage time ($t_{50}=130$ vs.\ 254 and 279) at comparable noise shift ($22.43\%$). Because $\ket{0}^{\otimes N}$ is a joint eigenstate of the longitudinal generators, broad longitudinal angles in Groups A and B merely imprint random relative phases onto the small transverse excitations, slightly retarding late-stage fine convergence rather than providing noise protection.

\begin{figure*}[t]
\includegraphics[width=\textwidth,clip]{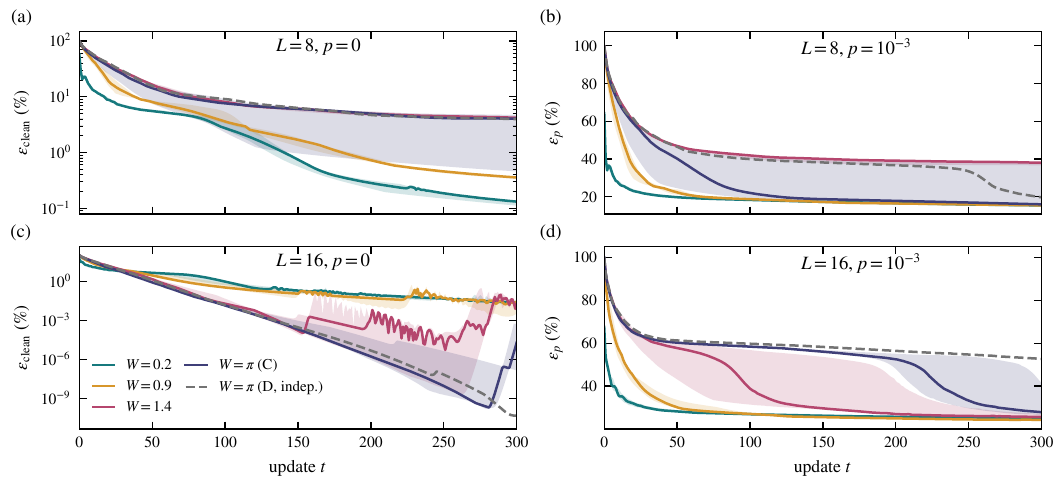}
\caption{Transverse-width scan at $N=8$ (seeds 0--4; medians and IQRs) with repeated layers: $W=0.2$ (Group A), 0.9, 1.4, and $\pi$ (Group C); the dashed line is the independent-layer Group D ($W=\pi$, median only). Panels: (a) $L=8$, $p=0$; (b) $L=8$, $p=10^{-3}$; (c) $L=16$, $p=0$; (d) $L=16$, $p=10^{-3}$. The late-time sawtooth oscillations in (c) occur at $\ecl\lesssim10^{-4}\%$ and reflect fixed-step ($\eta=0.03$) Adam overshooting near the exact minimum; they do not affect the noisy dynamics.}\label{fig:width}
\end{figure*}

Full-width longitudinal disorder and Floquet layer repetition are therefore unnecessary; the essential requirement shared by Groups A, B, and E is bounding the initial transverse generator amplitudes ($|\theta_\perp|\le0.2$). Figure~\ref{fig:width} corroborates this threshold with a single-parameter scan of the transverse width $W\in\{0.2,0.9,1.4,\pi\}$ at repeated layers, in which $W=0.2$ and $\pi$ correspond to seeds 0--4 of Groups A and C and the independent-layer Group D serves as a reference. In five-seed medians at $L=16$, $p=10^{-3}$, circuits with $W=0.2$ and $W=0.9$ bypass the high-noise plateau and cross $\ep=30\%$ at 30 and 51 updates, respectively. By contrast, $W=1.4$ stalls above $50\%$ error until update 84 (crossing $30\%$ at update 145), $W=\pi$ (Group C) remains on the plateau until update 213 (crossing $30\%$ at update 270), and the five-seed median of independent-layer Group D remains between $52.6\%$ and $61.2\%$ from update 50 to update 300. This dynamical crossover reflects the initial-state structure at $t=0$ listed in Table~\ref{tab:initial} ($L=N$, $p=0$, 50 seeds): across $N\in\{4,6,8,10\}$, $W=0.2$ circuits remain in an $N$-independent, low-entanglement, computational-basis-localized regime with half-chain entropy $S_{1/2}\approx0.12$--$0.19$~nats and inverse participation ratio $\mathrm{IPR}\approx0.39$--$0.79$ \cite{Sierant2025}. Conversely, $W=1.4$ already approaches volume-law delocalization, with $S_{1/2}$ within $0.05$~nats of the finite-size Page mean $S_{\mathrm{Page}}=\sum_{k=D+1}^{D^2}k^{-1}-(D-1)/(2D)$ ($D=2^{N/2}$) \cite{Page1993} and $\mathrm{IPR}$ near the Haar limit $2/(2^N+1)$.

\begin{table*}[t]
\caption{Initial-state and initial-gradient statistics at $t=0$ for $N\in\{4,6,8,10\}$ (50 seeds per condition). $S_{1/2}$ and IPR: medians at $L=N$, $p=0$, for transverse width $W$ with repeated layers and longitudinal angles in $[-\pi,\pi)$; Page: finite-size Page mean $\sum_{k=D+1}^{D^2}k^{-1}-(D-1)/(2D)$ with $D=2^{N/2}$; Haar IPR $2/(2^N+1)$. Var: ensemble variance of the middle-layer $R_x$ derivative of $E/N$ at $L=8$ for $W=0.2$ (Group A type) and full-width random angles (Group D type), at $p=0$ and $p=10^{-3}$.}\label{tab:initial}
\centering\scriptsize
\begin{tabular*}{\textwidth}{@{\extracolsep{\fill}}lcccccccccc@{}}
\toprule
 & \multicolumn{3}{c}{$S_{1/2}$ (nats)} & \multicolumn{3}{c}{IPR} & \multicolumn{2}{c}{$10^3\,\mathrm{Var}$, $p=0$} & \multicolumn{2}{c}{$10^3\,\mathrm{Var}$, $p=10^{-3}$}\\
\cmidrule(lr){2-4}\cmidrule(lr){5-7}\cmidrule(lr){8-9}\cmidrule(lr){10-11}
$N$ & $W=0.2$ & $W=1.4$ & Page & $W=0.2$ & $W=1.4$ & Haar & $W=0.2$ & Full & $W=0.2$ & Full\\
\midrule
4 & 0.124 & 0.880 & 0.922 & 0.785 & 0.120 & 0.118 & 17.2 & 17.0 & 11.6 & 10.2\\
6 & 0.184 & 1.569 & 1.589 & 0.548 & 0.0307 & 0.0308 & 8.06 & 4.30 & 5.51 & 2.10\\
8 & 0.152 & 2.230 & 2.275 & 0.481 & 0.0081 & 0.0078 & 3.70 & 0.651 & 2.51 & 0.269\\
10 & 0.188 & 2.917 & 2.966 & 0.388 & 0.0020 & 0.0020 & 3.25 & 0.132 & 2.18 & 0.0495\\
\bottomrule
\end{tabular*}
\end{table*}

Physically, this protection stems from finite-depth suppression of operator spreading near the Abelian longitudinal subgroup. Conjugating a single-site transverse operator by a longitudinal two-qubit gate generates only a nearest-neighbor diagonal tail,
\begin{equation}\label{eq:halo}
R_{zz}^{\dagger}(\theta)X_iR_{zz}(\theta)
=\cos\theta\,X_i-\sin\theta\,Y_iZ_{i+1}.
\end{equation}
Because this diagonal tail commutes with all subsequent longitudinal gates ($[Z_{i+1},Z_{i+1}Z_{i+2}]=0$), purely longitudinal evolution cannot expand the operator support further. Extending the spatial cascade requires another transverse rotation to convert $Z_{i+1}$ back into a noncommuting transverse operator, so each step of operator growth is suppressed by $\sin\theta_\perp\sim O(W)$. Thus, while longitudinal gates entangle existing superpositions, small transverse amplitudes bottleneck multistep operator delocalization.

Layer repetition plays a notable role only within the delocalized full-width regime: Group C reaches $\epsilon_\star=30\%$ 73 updates earlier than Group D ($t_{50}=270$ vs.\ 343), with similar leads at $\epsilon_\star=35\%$ (241 vs.\ 316) and $25\%$ (471 vs.\ 524), whereas confined Groups A and B show no such gap. By restricting $\bm\theta(0)$ to an $L$-fold smaller submanifold of identical layers, layer repetition provides a collective initialization prior that accelerates late-stage escape from the high-noise plateau analyzed in Sec.~\ref{sec:mechanism}. In short, initial transverse confinement selects the low-fragility optimization regime, whereas layer repetition acts only as a secondary accelerator within the fragile full-width regime.

\subsection{Robustness of the speed--fragility tradeoff}\label{sec:robust}

\begin{figure*}[t]
\includegraphics[width=\textwidth,clip]{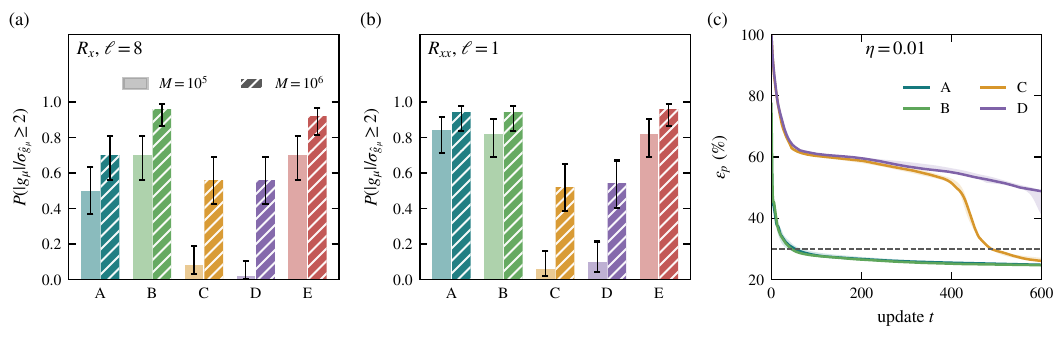}
\caption{(a),(b) Fraction of 50 initial seeds per group with resolved single-gate derivatives, $|g_\mu|/\sigma_{\widehat g_\mu}\geq2$ [Eqs.~\eqref{eq:shotg} and \eqref{eq:shotsigma}], for (a) a middle-layer ($\ell=8$) $R_x$ gate and (b) a first-layer $R_{xx}$ gate at $M=10^5$ (solid) and $10^6$ (hatched) shots per derivative; error bars denote Wilson 95\% intervals. (c) Noisy retraining of Groups A--D at learning rate $\eta=0.01$ (five seeds per group; medians and IQRs); dashed line, 30\% target.}\label{fig:shots}
\end{figure*}

At initialization ($t=0$), the dimensionless coherent gradient norm $\lVert\bm g_0\rVert=\lVert\nabla \Ezero\rVert/|\Egs|$ is 3.3--3.5 times larger in confined circuits ($1.23$ and $1.18$ for Groups A and B) than in full-width circuits ($0.355$ and $0.360$ for C and D). As shown in Fig.~\ref{fig:shots}, this contrast translates directly into single-gate derivative resolvability under parameter-shift measurements \cite{Sweke2020,Wierichs2022,Harrow2021,Kaminishi2026}. With $M=10^5$ shots per derivative, a middle-layer ($\ell=8$) $R_x$ derivative reaches a signal-to-noise ratio $\mathrm{SNR}\equiv|g_\mu|/\sigma_{\widehat g_\mu}\ge2$ in $50\%$, $70\%$, and $70\%$ of seeds for Groups A, B, and E versus $8\%$ and $2\%$ for C and D; for a first-layer $R_{xx}$ gate, the resolved fractions are $84\%$, $82\%$, and $82\%$ versus $6\%$ and $10\%$ (increasing to $70\%$--$96\%$ vs.\ $52\%$--$56\%$ at $M=10^6$). Consistently, the ensemble variance of the initial middle-layer $R_x$ derivative from $N=4$ to $N=10$ ($L=8$, $p=0$) drops by a factor of ${\sim}130$ for full-width circuits but only ${\sim}5$ for $W=0.2$, so that at $N=10$ the $W=0.2$ variance exceeds the full-width value by factors of $24.5$ ($p=0$) and $44.0$ ($p=10^{-3}$) at $L=8$ (Table~\ref{tab:initial}); for square circuits ($L=N=10$) the corresponding factors are $21.7$ and $53.0$. This rapid finite-size decay in full-width circuits reflects standard barren-plateau scaling in delocalized random circuits \cite{McClean2018,Holmes2022,OrtizMarrero2021}, which small-angle initialization mitigates \cite{Grant2019,Park2024}. Notably, larger individual initial derivatives in confined circuits do not produce faster early coherent descent at $N=8$; as discussed in Sec.~\ref{sec:separation}, full-width circuits there benefit from simultaneous noncommuting couplings across all bonds. Estimating all $P=640$ derivatives independently at $M=10^5$ would require $PM=6.4\times10^7$ shots per update.

The speed--fragility tradeoff is also insensitive to the optimizer step size. Retraining at a reduced learning rate $\eta=0.01$, shown in Fig.~\ref{fig:shots}(c), preserves the group ordering: Groups A and B cross $\ep=30\%$ in 5/5 seeds (median 53 and 45 updates; endpoints $24.79\%$ and $24.66\%$), Group C in 4/5 seeds ($t_{50}=488$; endpoint $26.06\%$), and Group D in 1/5 seeds (median endpoint $48.80\%$). Moreover, the early noise-shift peaks of Groups C and D reach $59.67\%$ and $60.33\%$, matching the $\eta=0.03$ peaks ($59.54\%$ and $59.39\%$). Lowering $\eta$ thus stretches the optimization timeline without altering the $60\%$ noise excursion or the plateau structure.

Finally, the early coherent-descent advantage of full-width circuits disappears upon increasing the system size to the $N=10$, $L=16$ noiseless reference case ($P=800<2^{11}-2$). Here, confined Groups A and B reach $\ecl\le5\%$ in 5/5 seeds (median 37 and 34 updates) while maintaining low fault susceptibility ($\chi/|\Egs|=334.6$ and $336.0$) and localized operator support ($\Rop=1.447$ and $1.446$). By contrast, only one of the ten full-width seeds (in Group C) reaches $\ecl\le5\%$ (at update 86, with $\chi/|\Egs|=620.3$ and $\Rop=2.23$); the remaining nine seeds stall at 600-update median coherent errors of $16.03\%$ (C) and $16.11\%$ (D) with high susceptibility ($\chi/|\Egs|=1029.2$ and $1014.6$) and fully delocalized support ($\Rop=2.92$). Rapid full-width coherent descent at $N=8$, $L=16$ is therefore an artifact of full-Hilbert-space overparameterization; in underparameterized larger systems, full-width circuits stall in delocalized, high-susceptibility regions even at $p=0$, whereas confined circuits remain trainable.

\section{Fault erasure and force-balance dynamics}\label{sec:mechanism}

\subsection{Phase-space trajectories and breakdown of uniform attenuation}

\begin{figure*}[t]
\includegraphics[width=\textwidth,clip]{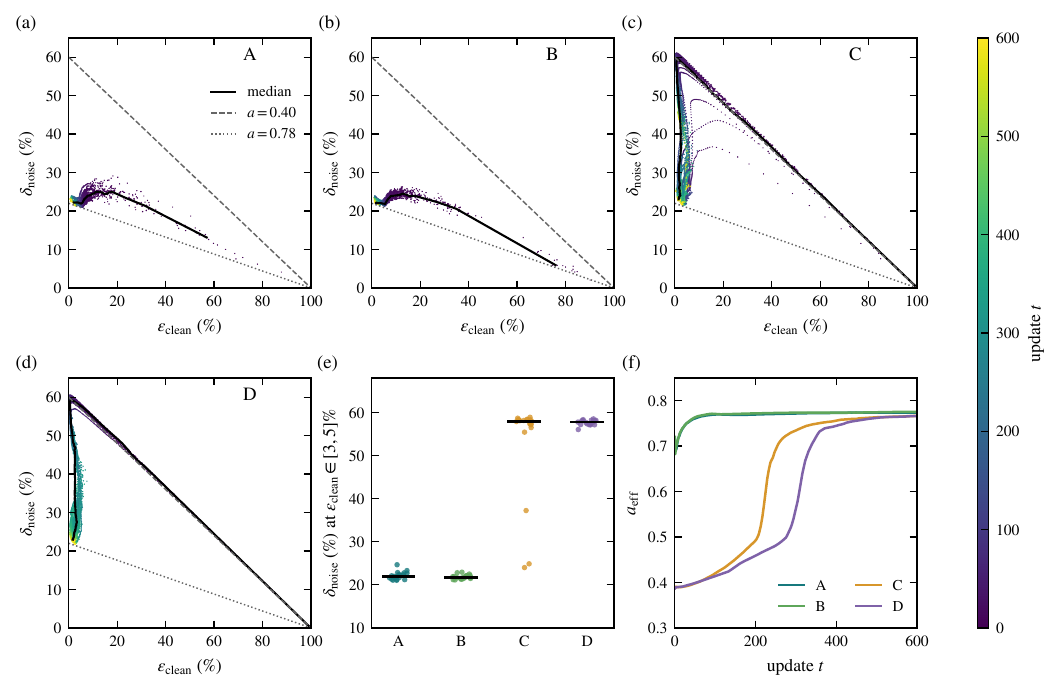}
\caption{(a)--(d) Trajectories of Groups A--D in the $(\ecl,\dn)$ plane over 600 updates (20 seeds per group). Points are colored by update $t$, black curves trace stepwise medians, and gray lines indicate the uniform-attenuation reference [Eq.~\eqref{eq:null}] at $a=0.40$ (dashed) and $a=0.78$ (dotted). (e) Noise shift $\dn$ upon first entering $\ecl\in[3\%,5\%]$ (points, individual seeds; bars, medians). (f) Median effective visibility $\aeff(t)$ across trajectory points with $\ecl<0.9$.}\label{fig:phase}
\end{figure*}

Figure~\ref{fig:phase} projects the 600-update trajectories onto the $(\ecl,\dn)$ plane and reveals two distinct optimization paths. Full-width circuits (Groups C and D) undergo rapid stage-I coherent descent, crossing $\ecl=5\%$ at 31 and 32 updates while their noise shift surges to median peaks of $59.54\%$ (C, update 49.5) and $59.39\%$ (D, update 58). Conversely, confined circuits (Groups A, B, and E) follow a low-noise path from $t=0$, with per-seed noise shifts peaking at $25.44\%$ (A) and $24.71\%$ (B) within the first ten updates before decaying smoothly. Once $\ecl$ becomes small, the full-width trajectories transition into stage II---a slow, nearly vertical descent along the $\dn$ axis at nearly constant coherent error. Within the first 300 updates, 18/20 (C) and 20/20 (D) seeds exhibit at least one iso-energetic denoising window, compared with only 5/20 (A), 2/20 (B), and 0/20 (E). By update 600, Groups A--D converge to comparable endpoints ($\ecl=1.89\%$, $1.71\%$, $1.59\%$, $1.60\%$ and $\dn=22.27\%$, $22.01\%$, $23.01\%$, $22.96\%$) from opposite directions: confined circuits via a direct low-noise path, and full-width circuits via a high-noise detour followed by prolonged iso-energetic recovery.

These stage-II trajectories violate the state-independent uniform-attenuation model of Eq.~\eqref{eq:null}. Fitting a single constant visibility $a$ to all 601 points of each seed yields large median root-mean-square residuals of $14.02$~pp (C) and $14.91$~pp (D), versus $0.99$~pp (A) and $0.93$~pp (B). Figure~\ref{fig:phase}(e) compares the groups at matched coherent error, upon first entering $\ecl\in[3\%,5\%]$, where Groups C and D incur $\dn=57.94\%$ and $57.73\%$ versus $21.87\%$ and $21.71\%$ for Groups A and B, representing a $36$~pp ($2.6$-fold) noise penalty. Consistently, the effective visibility $\aeff(t)$ in Fig.~\ref{fig:phase}(f) remains near $0.77$ throughout training for Groups A and B, whereas for Groups C and D it drops to $0.400$ and $0.399$ at update 50, recovers to $0.73$ and $0.56$ at update 300 (20-seed medians), and merges into the shared ${\sim}0.77$ band by update 600. Notably, the stage-I minimum $\aeff\approx0.40$ lies close to the zero-fault survival probability $(1-p)^K=0.358$, corresponding to the limit where every single-qubit fault completely erases the coherent energy. Thus, full-width circuits behave locally as strongly attenuating circuits during stage I, and the global uniform-attenuation model fails because optimization actively reshapes $\aeff$ from ${\sim}0.40$ to ${\sim}0.77$ during stage II.

\subsection{Energetic fault erasure versus spatial delocalization}

\begin{figure*}[t]
\includegraphics[width=\textwidth,clip]{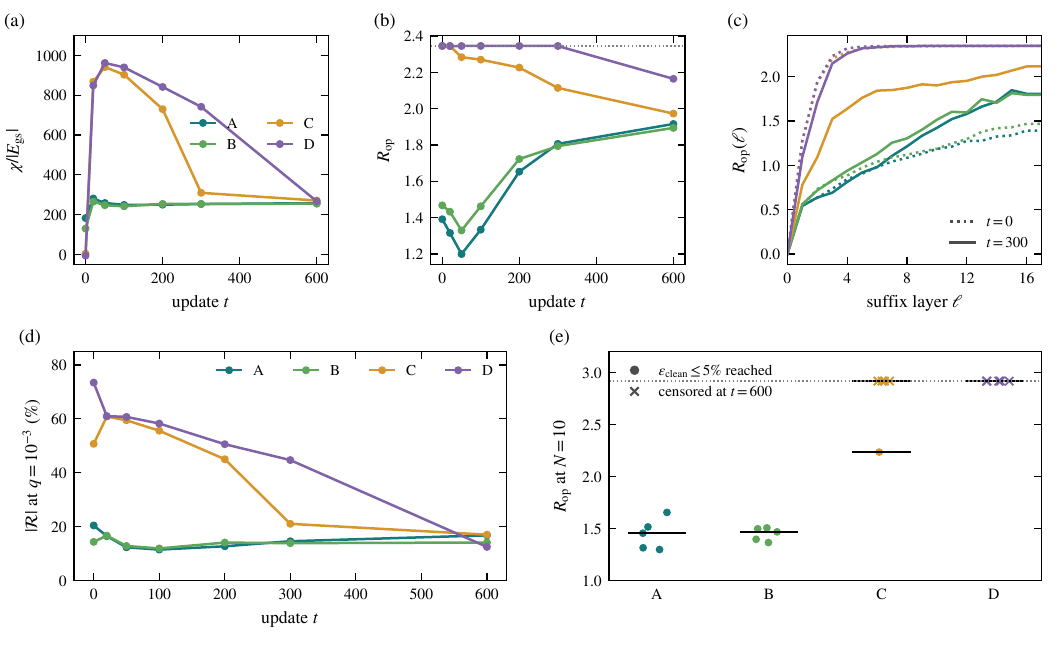}
\caption{Microscopic diagnostics across seven snapshots (seeds 0--4). (a) Median normalized susceptibility $\chi/|\Egs|$ estimated from $m=24$ sampled fault locations. (b) Five-seed mean operator radius $\Rop$; dotted line, complete-delocalization reference $R_{\mathrm{op}}^{\mathrm{unif}}$. (c) Layer-resolved profile $\Rop(\ell)$ at $t=0$ (dotted) and $t=300$ (solid). (d) Median relative residual $|\mathcal R|$ of the linear noise estimate at replay noise rate $q=10^{-3}$. (e) Endpoint $\Rop$ for the noiseless $N=10$ reference case: circles, seeds reaching $\ecl\leq5\%$; crosses, seeds censored at $t=600$; bars, subset medians; dotted line, $R_{\mathrm{op}}^{\mathrm{unif}}$.}\label{fig:micro}
\end{figure*}

\begin{figure*}[t]
\includegraphics[width=\textwidth,clip]{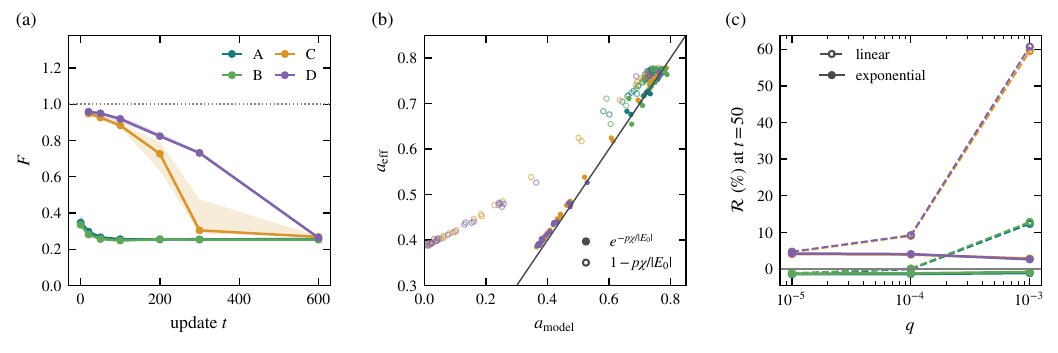}
\caption{(a) Median fault-erasure fraction $F$ (with IQRs) for snapshots with $\ecl<0.9$ (omitting $t=0$ for Groups C and D where $\Ezero\simeq0$); dotted line, $F=1$. (b) Exact visibility $\aeff$ at $p=10^{-3}$ versus the parameter-free exponential prediction $a_{\mathrm{model}}=e^{-p\chi/|\Ezero|}$ and the linear estimate $1-p\chi/|\Ezero|$ (open) across all 130 snapshot configurations with $\ecl<0.9$. (c) Median signed relative residual $\mathcal R$ of the linear (dashed, open) and exponential (solid, filled) predictions at $t=50$ versus replay noise rate $q$.}\label{fig:erasure}
\end{figure*}

Figure~\ref{fig:micro} tracks the energy susceptibility $\chi$ and operator radius $\Rop$ across the seven snapshots. At $t=0$, full-width circuits exhibit negligible net susceptibility [$\chi/|\Egs|=4.5$ (C) and $-5.4$ (D) vs.\ $183$ (A) and $130$ (B)] despite already having fully delocalized operator support: Fig.~\ref{fig:micro}(b) gives $\Rop=2.345$ and $2.346$, matching $R_{\mathrm{op}}^{\mathrm{unif}}$, versus $1.391$ and $1.467$ for A and B, and the layer-resolved profiles in Fig.~\ref{fig:micro}(c) show an injected fault spreading across the entire ring within four suffix layers. The same spatial dichotomy holds in the noiseless $N=10$ reference runs of Fig.~\ref{fig:micro}(e), where confined seeds converge at $\Rop\simeq1.45$ while unhit full-width seeds saturate at $R_{\mathrm{op}}^{\mathrm{unif}}=2.92$. The vanishing initial susceptibility of full-width circuits follows directly from $\chi=KF|\Ezero|$ in Eq.~\eqref{eq:F}: because random initial states sit near $\Ezero\simeq0$ ($\ecl=99.6\%$ and $100.4\%$), individual fault responses $\Delta E_s$ fluctuate symmetrically around zero and cancel in the sum. Figure~\ref{fig:erasure}(a) shows that, as stage-I optimization lowers $\Ezero$, $\chi$ grows linearly with $|\Ezero|$ at a nearly constant, near-unity fault-erasure fraction: $F=0.948$ (C) and $0.958$ (D) at update 20, and $0.927$ and $0.950$ at update 50. Consequently, $\chi/|\Egs|$ surges to 868 and 849 at update 20 and to $941.5$ and $962.0$ at update 50---approaching the white-noise limit $K(1-\ecl)\simeq1016$ as the fraction of fault locations with $\Delta E_s>0$ reaches unity. By contrast, confined circuits maintain $F\approx0.35$ and $0.33$ at $t=0$ and $F=0.25$--$0.27$ from update 50 onward, keeping $\chi/|\Egs|$ bounded between 242 and 281 across $t=20$--$600$. Stage-I fragility in full-width circuits is therefore energetic rather than newly generated spatial spreading: the circuits are spatially scrambling from $t=0$ ($\Rop=2.345$), and coherent energy descent simply builds the negative energy $\Ezero$ that faults subsequently erase.

Spatial operator support relaxes on a slower time scale than energetic susceptibility. In confined Groups A and B, $\Rop$ initially contracts during early descent (from $1.391$ and $1.467$ at $t=0$ to $1.199$ and $1.329$ at $t=50$) before expanding to $1.806$ and $1.794$ at $t=300$ and $1.916$ and $1.894$ at $t=600$ as ground-state correlations develop. In full-width Groups C and D, stage-II escape from the plateau is accompanied by spatial contraction: by $t=300$, Group C has exited the plateau with $\Rop$ reduced to $2.114$, whereas Group D remains on the plateau at $\Rop=2.345$, as seen in Figs.~\ref{fig:micro}(b) and \ref{fig:micro}(c). Crucially, the fault-erasure fraction $F$ drops before $\Rop$ contracts: in Group D (seeds 0--4), $F$ declines from $0.92$ to $0.73$ between updates 100 and 300 while $\Rop$ remains saturated at $2.345$. For these five seeds the exact visibility at update 300 is $\aeff=0.48$, consistent with $e^{-pKF}=0.47$; the larger 20-seed median in Fig.~\ref{fig:phase}(f), $0.56$, includes seeds that have already left the plateau. This separation reflects the distinct observables in Eq.~\eqref{eq:commutator}: $\Rop$ measures state-independent operator support, whereas $\Delta E_{s,Q}=\bra{\psi_0}W_{s,Q}^\dagger[H,W_{s,Q}]\ket{\psi_0}$ depends on the alignment between the propagated fault $W_{s,Q}$ and the state $\ket{\psi_0}$. Optimization thus reduces energetic fragility first by reorienting propagated faults relative to $\ket{\psi_0}$, and only later contracts the operator light cone. By update 600, all four groups converge to a common fault-erasure fraction $F=0.25$--$0.27$ and susceptibility $\chi/|\Egs|=255$--$271$, whereas their mean radii $\Rop$ are $1.916$ (A, IQR $1.91$--$2.00$), $1.894$ (B, $1.85$--$1.93$), $1.973$ (C, $1.92$--$2.05$), and $2.164$ (D, $2.06$--$2.22$). Group C has contracted back into the confined-group range, whereas Group D---which exited the plateau last---retains a larger but still rapidly shrinking radius (dropping from $2.345$ to $2.164$ between $t=300$ and $600$, mirroring Group C's contraction ${\sim}100$ updates earlier). We refer to this delayed spatial contraction at matched noisy energy and matched $F$ as finite-budget trajectory path memory.

Quantitatively, assuming faults at different locations act independently so that each reduces the surviving coherent energy by a factor $(1-f_s)$, the noisy energy factorizes as $E_p=\Ezero\prod_{s=1}^K(1-pf_s)$. Expanding to leading order in $pf_s$ yields the parameter-free exponential visibility model
\begin{equation}\label{eq:expvis}
\aeff\simeq e^{-pKF},\qquad
\dn\simeq(1-\ecl)\left(1-e^{-pKF}\right),
\end{equation}
which reduces to the uniform attenuation of Eq.~\eqref{eq:null} for constant $F$ and to the linear zero-noise estimate $\dn\simeq p\chi/|\Egs|$ at first order in $pKF$. As shown in Fig.~\ref{fig:erasure}(b), across all 130 snapshot configurations with $\ecl<0.9$, Eq.~\eqref{eq:expvis} predicts the exact density-matrix visibility $\aeff$ with a median absolute error of $0.006$ and a maximum error of $0.046$, whereas the linear estimate $1-p\chi/|\Ezero|$ deviates by up to $0.40$. In replay scans across noise rates $q$, measured via the signed relative residual $\mathcal R=(\dn^{\mathrm{model}}-\dn)/\dn$, the linear approximation overshoots the exact noise shift at $q=10^{-3}$ ($Kq=1.028$) by $+59.46\%$ (C) and $+60.72\%$ (D) at update 50, as plotted in Fig.~\ref{fig:micro}(d), where $pKF\simeq1$ makes multiplicative fault compounding essential. Figure~\ref{fig:erasure}(c) shows that the exponential model reduces these residuals to $2.8\%$ and $2.6\%$, lowering the median absolute residual across all 130 configurations at $q=10^{-3}$ from $17\%$ to $2\%$. The residual $2\%$--$4\%$ offset is nearly independent of $q$ across $q\in\{10^{-5},10^{-4},10^{-3}\}$ (e.g., $4.1\%$, $3.9\%$, and $2.8\%$ for Group C), consistent with the $1\%$--$6\%$ finite-population sampling error of the 24-location estimator $\widehat{\chi}$. Finally, substituting the shared 600-update value $F\simeq0.26$ into Eq.~\eqref{eq:expvis} yields $\dn\simeq1-e^{-1.028\times0.26}\simeq23\%$, quantitatively accounting for the common late-time noisy energy band.

\subsection{Gradient decomposition and force-balance plateau}

Differentiating the visibility relation $\dn=(1-\aeff)(1-\ecl)$, which follows from Eq.~\eqref{eq:aeff}, decomposes the noise-induced gradient into a uniform-attenuation component opposing $\bm g_0$ and a visibility-gradient force $\bm b$, which Eq.~\eqref{eq:expvis} links directly to $\nabla_{\bm\theta}F$:
\begin{align}\label{eq:split}
\bm g_\delta&=-(1-\aeff)\bm g_0-(1-\ecl)\nabla_{\bm\theta}\aeff\nonumber\\
&\simeq-(1-a)\bm g_0+(1-\ecl)apK\nabla_{\bm\theta}F.
\end{align}
In the second line, $\aeff$ is replaced by the model visibility $a\equiv e^{-pKF}$ of Eq.~\eqref{eq:expvis}, so that $\nabla_{\bm\theta}a=-apK\nabla_{\bm\theta}F$. Here, the visibility-gradient force $\bm b\equiv\bm g_\delta+(1-\aeff)\bm g_0=-(1-\ecl)\nabla_{\bm\theta}\aeff$ drives parameters toward higher visibility (lower $F$), with relative magnitude
\begin{equation}\label{eq:b}
\frac{\lVert\bm b\rVert}{\lVert\bm g_0\rVert}
=\left[r^2+(1-\aeff)^2+2r(1-\aeff)\cos\phi\right]^{1/2}.
\end{equation}
Because Adam's step normalization $\widehat m_t/(\sqrt{\widehat v_t}+\epsilon_{\mathrm{Adam}})$ is invariant under positive scalar rescaling (up to $\epsilon_{\mathrm{Adam}}$), an attenuation-dominated total gradient $\nabla_{\bm\theta}\ep\simeq \aeff\bm g_0$ (where $\lVert\bm b\rVert\ll\lVert\bm g_0\rVert$ and $\cos\psi\simeq1$) leaves the optimization trajectory nearly identical to noiseless descent.

\begin{figure*}[t]
\includegraphics[width=\textwidth,clip]{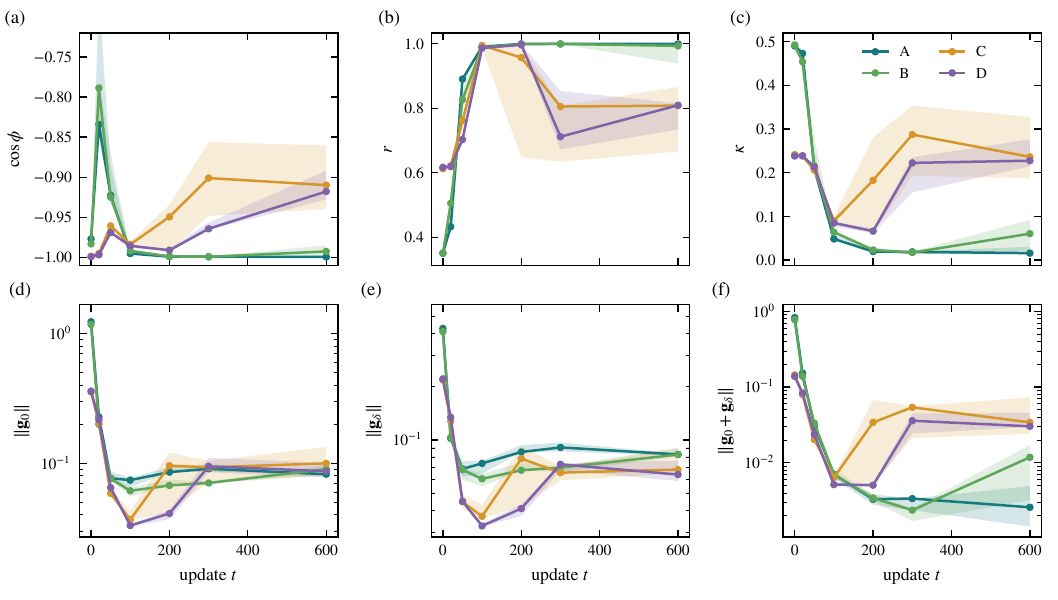}
\caption{Gradient geometry across seven snapshots (seeds 0--4; medians and IQRs). (a)--(c) Cosine alignment $\cos\phi$, norm ratio $r$, and cancellation factor $\kappa$ . (d)--(f) Gradient norms $\lVert\bm g_0\rVert$, $\lVert\bm g_\delta\rVert$, and $\lVert\bm g_0+\bm g_\delta\rVert$ on logarithmic scales.}\label{fig:grad}
\end{figure*}

\begin{table*}[t]
\caption{Stage-I gradient geometry (medians over seeds 0--4). Values of $F$ and $\aeff$ are omitted for Groups C and D at $t=0$ ($\ecl\ge0.9$), where $\Ezero\simeq0$ renders the normalized ratios ill-conditioned; $\lVert\bm b\rVert/\lVert\bm g_0\rVert$ is evaluated via Eq.~\eqref{eq:b}.}\label{tab:grad}
\centering\scriptsize
\begin{tabular*}{\textwidth}{@{\extracolsep{\fill}}lcccccccc@{}}
\toprule
Group & $t$ & $\ecl$ (\%) & $F$ & $1-\aeff$ & $r$ & $\cos\phi$ & $\cos\psi$ & $\lVert\bm b\rVert/\lVert\bm g_0\rVert$ \\
\midrule
A & 0 & 53.8 & 0.35 & 0.30 & 0.35 & $-0.977$ & 0.993 & 0.10 \\
A & 20 & 7.8 & 0.30 & 0.26 & 0.43 & $-0.834$ & 0.936 & 0.27 \\
A & 50 & 5.8 & 0.27 & 0.24 & 0.89 & $-0.923$ & 0.457 & 0.68 \\
B & 0 & 68.1 & 0.33 & 0.29 & 0.35 & $-0.983$ & 0.995 & 0.08 \\
B & 20 & 7.9 & 0.28 & 0.25 & 0.51 & $-0.789$ & 0.887 & 0.35 \\
B & 50 & 5.9 & 0.26 & 0.23 & 0.83 & $-0.925$ & 0.599 & 0.61 \\
C & 0 & 99.6 & -- & -- & 0.61 & $-0.999$ & 0.997 & -- \\
C & 20 & 10.4 & 0.95 & 0.61 & 0.62 & $-0.995$ & 0.988 & 0.06 \\
C & 50 & 1.15 & 0.93 & 0.60 & 0.76 & $-0.961$ & 0.785 & 0.25 \\
D & 0 & 100.4 & -- & -- & 0.62 & $-0.999$ & 0.998 & -- \\
D & 20 & 13.9 & 0.96 & 0.61 & 0.62 & $-0.997$ & 0.992 & 0.05 \\
D & 50 & 1.33 & 0.95 & 0.61 & 0.70 & $-0.969$ & 0.878 & 0.19 \\
\bottomrule
\end{tabular*}
\end{table*}

Table~\ref{tab:grad} shows that full-width circuits operate in this pure-attenuation regime throughout early stage I. Although $\bm g_\delta$ is large at $t=0$ ($\lVert\bm g_\delta\rVert=0.22$, $r=0.61$), it is nearly antiparallel to $\bm g_0$ ($\cos\phi\le-0.995$) with magnitude matching the uniform-attenuation factor ($r=0.62$ vs.\ $1-\aeff=0.61$ at $t=20$). Because the circuit is already maximally scrambling ($\Rop=2.345$, $F\simeq0.95$), the local visibility landscape is flat ($\lVert\bm b\rVert/\lVert\bm g_0\rVert=0.05$--$0.06$ at $t=20$). The total gradient thus remains collinear with $\bm g_0$ ($\cos\psi\ge0.988$ through $t=20$), explaining why Adam's scale-invariant updates drive Groups C and D across $\ecl=5\%$ in 31--32 updates regardless of noise. Braking sets in only around $t=50$, when $\lVert\bm b\rVert/\lVert\bm g_0\rVert$ rises to $0.19$--$0.25$ and $r$ reaches $0.70$--$0.76$, slightly delaying the $\ecl=1\%$ crossing from 49.5--51 updates (at $p=0$) to 55--56 updates. Conversely, confined circuits (Groups A and B) start in a localized regime ($F\simeq0.3$) where $\nabla_{\bm\theta}F$ is appreciable from the outset: by $t=20$, $r=0.43$--$0.51$ already exceeds $1-\aeff=0.25$--$0.26$, $\lVert\bm b\rVert/\lVert\bm g_0\rVert=0.27$--$0.35$ is comparable to the attenuation term, and $\cos\phi=-0.83$ to $-0.79$. By $t=50$, $\bm b$ rotates the net gradient by $53^\circ$--$63^\circ$ away from $\bm g_0$ ($\cos\psi=0.46$--$0.60$), retarding the $\ecl=5\%$ replay crossing from ${\sim}42$ to 104 updates as the optimizer steers parameters to preserve low $F$.

Figure~\ref{fig:grad} shows that by update 100 all groups enter a regime of strong first-order force cancellation, characterized by $\cos\phi=-0.995$ (A), $-0.992$ (B), $-0.984$ (C), and $-0.986$ (D), norm ratios $r\simeq1$, and cancellation factors $\kappa=0.048$--$0.089$. Setting $\bm g_0+\bm g_\delta=\bm{0}$ in Eq.~\eqref{eq:split} yields the exact force-balance condition $\aeff\bm g_0=(1-\ecl)\nabla_{\bm\theta}\aeff=-\bm b$, or within the exponential model,
\begin{equation}\label{eq:balance}
\bm g_0\simeq-(1-\ecl)pK\nabla_{\bm\theta}F.
\end{equation}
Thus, optimization stalls on a quasi-stationary plateau where further coherent energy descent requires increasing the fault-erasure fraction $F$. For Groups C and D, this balance locks the circuit onto a high-noise plateau at $\ep\simeq58\%$--$60\%$ ($F\simeq0.9$). Unlike standard or noise-induced barren plateaus \cite{McClean2018,Wang2021} where individual gradients vanish, both $\bm g_0$ and $\bm g_\delta$ remain finite ($\lVert\bm g_0\rVert=0.0367$ and $0.0330$ at $t=100$), but their vector sum cancels to $\lVert\bm g_0+\bm g_\delta\rVert=0.0064$ and $0.0052$. Operationally, because hardware measurements access only the net gradient $\nabla_{\bm\theta}\ep=\bm g_0+\bm g_\delta$, this cancellation imposes a severe shot overhead: resolving a typical root-mean-square derivative component $|g_\mu|\sim(|\Egs|/N)\lVert\bm g_0+\bm g_\delta\rVert/\sqrt{P}\simeq3.8\times10^{-4}$ at $\mathrm{SNR}\ge2$ in the strongly depolarized regime ($\aeff\simeq0.4$, where $\sigma_{\widehat g_\mu}\simeq3/\sqrt{M}$ from Eq.~\eqref{eq:shotsigma}) requires $M\simeq2.5\times10^8$ shots per derivative, or $PM\sim1.6\times10^{11}$ shots per update. In exact-gradient simulation, Adam's moment normalization rescales the small uncancelled residual component, enabling slow stage-II escape along nearly iso-energetic directions that reduce $F$ (from $0.88$ to $0.30$ in Group C between $t=100$ and $300$ at $\ecl\in[0.7\%,1.7\%]$). For $t\ge300$, $r$ relaxes to $0.71$--$0.81$ and $\kappa$ to $0.22$--$0.29$ ($\lVert\bm g_0+\bm g_\delta\rVert=0.036$--$0.054$), sustaining stage-II denoising, while Groups A and B settle into a tight low-noise force balance ($\cos\phi\le-0.992$, $\kappa=0.016$--$0.060$ at $t=600$).

\section{Iso-energetic tangent projection}\label{sec:projection}

\subsection{Constrained noise-descent experiments}

\begin{figure*}[t]
\includegraphics[width=\textwidth,clip]{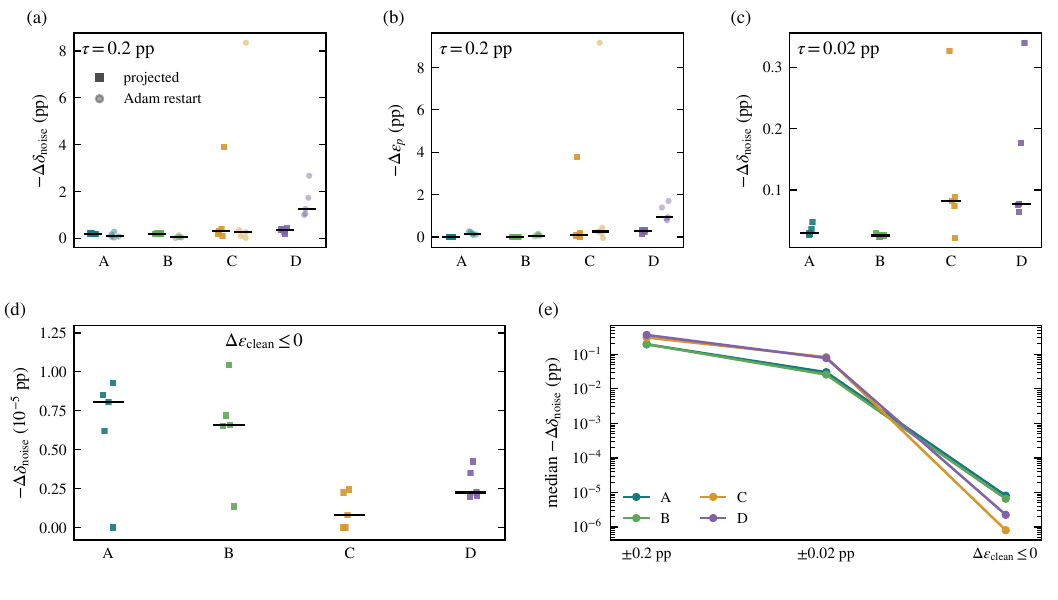}
\caption{Iso-energetic tangent projection from $t=300$ snapshots (five seeds per group). (a) Noise-shift reduction $-\Delta\dn$ in the tolerant tier ($\tau=0.2$~pp): squares, projected branches; circles, unconstrained Adam restarts; bars, medians. (b) Net noisy-error reduction $-\Delta\ep$ in the tolerant tier. (c) Projected $-\Delta\dn$ in the tight tier ($\tau=0.02$~pp). (d) Projected $-\Delta\dn$ under strict nondegradation ($\Delta\ecl\le10^{-12}$, in units of $10^{-5}$~pp). (e) Median projected $-\Delta\dn$ across tiers, showing a five-order-of-magnitude drop governed by the $\sqrt{\tau}$ curvature bound and discrete line-search floor.}\label{fig:repair}
\end{figure*}

Because stage-II optimization removes stage-I fragility at nearly constant $\ecl$, we test whether accumulated noise can be stripped post hoc by stepping along the tangent direction of the constant-$\ecl$ manifold defined in Eq.~\eqref{eq:d}. Figure~\ref{fig:repair} shows the results of applying the 50-step projection protocol of Sec.~\ref{sec:probes} to the $t=300$ snapshots, with analogous tests performed at $t=100$. Each branch consumes 100 full-gradient evaluations, matched to the unconstrained Adam control, plus a median of 103--431 (tolerant), 439--608 (tight), and 765--806 (strict) line-search energy evaluations at $t=300$ (versus 3 for Adam restarts).

In the tolerant tier ($\tau=0.2$~pp), projected branches at $t=300$ reduce $\dn$ by medians of $0.19$, $0.20$, $0.31$, and $0.36$~pp for Groups A--D. For Groups A and B, however, this reduction is offset by coherent-error drift to the upper tolerance boundary ($\Delta\ecl\approx+0.20$~pp), yielding slightly negative net error reductions ($-\Delta\ep=-0.0053$ and $-0.0046$~pp). Even in Groups C and D, the net projected gains ($0.105$ and $0.299$~pp) fall short of unconstrained Adam restarts ($0.163$, $0.070$, $0.260$, and $0.939$~pp for A--D), which jointly adjust $\ecl$ and $\dn$. Tightening the coherence tolerance tenfold to $\tau=0.02$~pp reduces the projected noise-shift gains to $0.030$, $0.026$, $0.083$, and $0.078$~pp. Under strict coherent nondegradation ($\Delta\ecl\le10^{-12}$), the gains collapse by five orders of magnitude to $(8.08,6.57,0.81,2.25)\times10^{-6}$~pp, the collapse summarized in Fig.~\ref{fig:repair}(e), as step acceptance drops from $91\%$ (tolerant) and $76\%$ (tight) to $12\%$ (strict, where $88\%$ of steps fail all 16 backtracking trials). Branches from $t=100$ exhibit identical scaling, with median $-\Delta\dn$ dropping from $0.108$--$0.287$~pp (tolerant) and $0.055$--$0.150$~pp (tight) to $(3.35\text{--}5.39)\times10^{-6}$~pp (strict), compared with tens of percentage points removed during unconstrained stage-II training.

\subsection{Geometry of the tangent step}\label{sec:tangent}

This failure of linear tangent projection follows from both first-order gradient alignment and second-order level-set curvature. For $\lVert\bm g_0\rVert\neq0$, the projected direction $\bm d$ of Eq.~\eqref{eq:d} satisfies
\begin{equation}\label{eq:tangent}
\bm g_0^\top\bm d=0,\quad
\bm g_\delta^\top\bm d=-\lVert\bm d\rVert^2,\quad
\lVert\bm d\rVert=\lVert\bm g_\delta\rVert|\sin\phi|.
\end{equation}
While $\bm d$ is a valid first-order descent direction for $\dn$ with directional derivative $-\lVert\bm g_\delta\rVert^2\sin^2\phi$, its norm is suppressed near force balance ($\phi\to\pi$). At $t=300$, median $|\sin\phi|$ is only $0.037$ (A) and $0.034$ (B) versus $0.43$ (C) and $0.26$ (D) (and $0.096$--$0.18$ across groups at $t=100$), leaving only $3\%$--$4\%$ of $\lVert\bm g_\delta\rVert$ accessible to tangent steps in the confined groups.

Furthermore, a straight Euclidean step $\alpha\bm d$ departs from the constant-$\ecl$ manifold at second order:
\begin{equation}\label{eq:curvature}
\Delta\ecl=\tfrac12\alpha^2\mu_d+O(\alpha^3),\qquad
\mu_d=\bm d^\top H_0\bm d,
\end{equation}
where $H_0=\nabla_{\bm\theta}^2\ecl$. For positive directional curvature ($\mu_d>0$), satisfying $\Delta\ecl\le\tau$ restricts the step size to $\alpha_{\max}\simeq(2\tau/\mu_d)^{1/2}$. Shrinking the allowance from the tolerant tier ($\tau=0.2$~pp, i.e., $2\times10^{-3}$ in units of $\ecl$) to the strict margin $10^{-12}$ reduces $\alpha_{\max}$ by a factor of $(10^{-12}/2\times10^{-3})^{1/2}\approx2.2\times10^{-5}$, pushing the admissible step down to the discrete backtracking floor $\alpha_{\min}=2^{-15}\approx3.1\times10^{-5}$. Once the first few minimal steps exhaust the $10^{-12}$ margin, all subsequent trials are rejected, explaining the five-decade collapse in cumulative gain (whereas $\mu_d<0$ would decrease $\ecl$ and permit acceptance). Effective iso-energetic denoising near a curved level set thus requires nonlinear manifold retraction or second-order curvature corrections, which unconstrained Adam implicitly approximates by alternating small coherent-error excursions with restorative gradient steps.

\section{Discussion and conclusion}\label{sec:conclusion}

We have followed complete trajectories of noisy variational training for a periodic Heisenberg chain under local gate depolarization. In the coherently exact regime ($N=8$, $L=16$), fast early coherent descent and fast noisy convergence work against each other. Full-width initializations lower the coherent error fastest. They also build up the largest noise shift and need an order of magnitude more updates to reach target noisy errors. Circuits with confined transverse generators descend more slowly in coherent energy but converge far earlier in noisy energy. Among the priors of MBL-inspired initialization, bounding the initial transverse generator amplitudes suffices for this low-fragility behavior. Full-width longitudinal disorder and Floquet layer repetition are unnecessary. Within 600 updates, all groups reach the same noisy-error band and the same fault-erasure fraction $F\simeq0.26$. The operator radius, however, contracts more slowly and still differs between groups. Group D stays longest on the plateau and keeps the widest operator support. We refer to this delayed contraction as finite-budget trajectory path memory.

We find that three relations explain these observations. The first relation follows from treating faults as independent and gives $E_p\simeq\Ezero e^{-pKF}$ with no free parameter. This relation reproduces the exact effective visibility at all analyzed snapshots. The noise shift is thus controlled by the product of fault exposure $pK$ and the mean erased fraction $F$. The second relation links visibility to the training dynamics. It splits the noise-induced gradient into a uniform attenuation of $\bm g_0$ and a visibility-gradient force $\bm b\propto-\nabla_{\bm\theta}F$. Full-width circuits start with saturated operator support, and their $F$ sits near unity once coherent energy appears. Their $\bm b$ is weak, so noise nearly uniformly rescales $\bm g_0$. Scale-invariant Adam updates absorb this rescaling, and coherent descent keeps its noiseless rate. Confined circuits start at low $F$. Their $\bm b$ is appreciable from the outset and steers the trajectory toward low fault sensitivity. The third relation defines the force-balance plateau. Setting $\bm g_0+\bm g_\delta=\bm0$ gives $\bm g_0\simeq-(1-\ecl)pK\nabla_{\bm\theta}F$. On this plateau, further coherent descent must raise $F$. Both gradients remain finite. These finite gradients separate the force-balance plateau from standard and noise-induced barren plateaus, in which the gradients themselves vanish. The same balance also limits post hoc tangent repair. Near force balance the tangential drive scales as $\sin^2\phi$, and the curvature of the constant-$\ecl$ manifold bounds the size of straight tangent steps.

We combine these relations into a single picture, in which fragility in these circuits is an energetic response of the current state. The operator radius $\Rop$ measures how far a propagated fault spreads. The fault-erasure fraction $F$ measures how much coherent energy that fault removes from $\ket{\psi_0}$ through the commutator expectation. Our data show that the two quantities can decouple. Full-width circuits are spatially scrambling at $t=0$, while their susceptibility stays negligible until coherent energy appears. Later in training, Group D lowers $F$ from 0.92 to 0.73 while $\Rop$ stays saturated. Spatial structure enters through the initial condition. Longitudinal gates commute with each other. When a longitudinal gate acts on a transverse operator, it adds only a nearest-neighbor diagonal tail, and further growth needs another transverse rotation. Small transverse amplitudes therefore keep the initial operator support localized and start the trajectory at low $F$. Initial delocalization places the circuit near $F\simeq1$ as soon as coherent energy builds up. Initialization selects the regime, and $F$ governs the dynamics within it.

Several parts of this picture remain open, because our simulations use exact gradients, a single depolarizing rate, unital noise, and an overparameterized eight-spin chain. Each of these choices raises a specific question. Doubling the depth from $L=8$ to $L=16$ makes the circuit coherently exact and also doubles $pK$, and our data cannot separate these two effects. The stage-I minimum $\aeff\approx0.40$ lies close to the zero-fault survival probability $(1-p)^K=0.358$. We therefore expect the plateau visibility to follow $e^{-pK}$ and the plateau to fade as $pK$ falls well below unity. A scan of $p$ at fixed $L=16$ would test this expectation. The unretarded stage-I descent also depends on the optimizer. Under plain gradient descent, uniform attenuation would slow coherent descent by the factor $\aeff$, and the coherent-speed advantage of full-width circuits should shrink.

System size raises a sharper question. At $N=10$ the circuit is underparameterized, and nine of ten full-width seeds stall even without noise. We conjecture that barren-plateau gradient decay and fault erasure reinforce each other as $N$ grows at fixed depth. The transient force-balance plateau would then become a lasting arrest for delocalized initializations. Finite-shot measurement adds stochastic gradient noise, whose relative size grows as $\bm g_0$ and $\bm g_\delta$ cancel. Moderate sampling noise could push the trajectory off the force-balance manifold and change the escape rate, much as thermal noise changes escape from a metastable state. Nonunital channels such as amplitude damping bias the state toward a polarized reference. Faults would then stop driving the energy toward zero, and the force-balance condition should acquire a directional drift term. Finally, because $F$ requires only zero-noise single-fault energy responses, it is a candidate for screening ans\"atze before noisy training.

\section*{Acknowledgments}
This work is supported by the National Natural Science Foundation of China (Grant No. 11975175). 

The authors have no conflicts to disclose.

\section*{Data Availability}
There are no publicly available research data or software supporting this manuscript. Requests for further information or data should be sent to the authors.

\bibliography{references_main}

\end{document}